\documentclass[10pt, conference, letterpaper]{IEEEtran}

\IEEEoverridecommandlockouts

\usepackage{amssymb}
\usepackage{graphicx,psfrag,url, amsmath,color,cite}
\usepackage[small,bf]{caption}
\usepackage[font=footnotesize]{subfig}
\usepackage{stfloats}
\usepackage{graphicx}
\usepackage[countmax]{subfloat}
\usepackage{paralist}
\usepackage{color}
\usepackage[english]{babel}
\usepackage[lined,vlined,ruled,linesnumbered]{algorithm2e}
\usepackage{multirow}
\usepackage{amsthm}

\newtheorem{theorem}{Theorem}
\newtheorem{lemma}{Lemma}

\newtheorem{definition}{Definition}
\newtheorem{proposition}{Proposition}

\newtheorem*{game}{Game}

\ifodd 121
\newcommand{\rev}[1]{{#1}}
\newcommand{\com}[1]{\textbf{\color{red} (Comment: #1)}}
\newcommand{\comlin}[1]{\textbf{\color{magenta} (Lin Comment: #1)}}
\else
\newcommand{\rev}[1]{{#1}}
\newcommand{\com}[1]{}
\newcommand{\comlin}[1]{}
\fi

\def\K{\mathcal{K}}
\def\Ks{\K_{\textsc{s}}}

\def\Ka{\K_{\textsc{a}}}
\def\Ku{\K_{\textsc{u}}}
\def\M{\mathcal{M}}

\def\L{\textsc{l}}
\def\B{\textsc{b}}

\begin{document}

\title {A Game-Theoretic Analysis of User Behaviors in Crowdsourced Wireless Community Networks}

\author{\IEEEauthorblockN{Qian Ma$^\ast$, Lin Gao$^\ast$, Ya-Feng Liu$^\dag$, and Jianwei Huang$^\ast$}
\IEEEauthorblockA{$^\ast$Dept. of Information Engineering, The Chinese University of Hong Kong, Hong Kong, China.\\
$^\dag$LSEC, AMSS, Chinese Academy of Sciences, Beijing, China.}
\thanks{\rev{This work is supported by the General Research Funds (Project Number CUHK 412713 and 14202814) established under the University Grant Committee of the Hong Kong Special Administrative Region, China.}}}

\maketitle

\begin{abstract}
A crowdsourced wireless community network can effectively alleviate the limited coverage issue of Wi-Fi access points (APs), by encouraging individuals (users) to share their private residential Wi-Fi APs with each other.
This paper presents the first study on the users' joint membership selection and network access problem in such a network.
Specifically, we formulate the problem as a two-stage dynamic game:
Stage I corresponds to a membership selection game, in which each user chooses his membership type;
Stage II corresponds to a set of network access games, in each of which each user decides his WiFi connection time on the AP at his current location.
We analyze the Subgame Perfect Equilibrium (SPE) systematically, and study whether and how best response dynamics can reach the equilibrium.
Through numerical studies, we further explore how the equilibrium changes with the users' mobility patterns and network access evaluations.
We show that a user with a more popular home location, a smaller travel time, or a smaller network access evaluation is more likely to choose a specific type of membership called Bill.
We further demonstrate how the network operator can optimize its pricing and incentive~mechanism based on the game equilibrium analysis in this work.
\end{abstract}

\section{Introduction}\label{sec:intro}

\subsection{Background and Motivation}\label{sec:moti}

The global mobile data traffic is growing explosively in recent years, with an anticipated annual growth rate of $61\%$ from 2013 to 2018 \cite{demand}.
The global cellular network capacity, however, grows much slower than the mobile data traffic.
To fill in such a gap, the Wi-Fi network is playing an increasingly important role in carrying the mobile data traffic.\footnote{According to Cisco's report \cite{demand}, around $45\%$ of the global mobile data traffic was offloaded to the fixed network through Wi-Fi or femtocell in 2013.}
The fast growth of Wi-Fi technology is due to several factors including the low cost of a Wi-Fi access point (AP), simple installation, easy management, and high data rate \cite{WiFiSpeed}.
However, the large-scale deployment of Wi-Fi networks is often restricted by the limited coverage of a single Wi-Fi AP (typically tens of meters indoors and hundreds of meters outdoors \cite{WiFiCoverage}), which is much smaller than the coverage of a cellular tower.
Hence, it is expensive for a single network operator to deploy enough Wi-Fi APs to cover an entire city or nation.

The crowdsourced wireless community network comes out as a promising solution to enlarge the Wi-Fi coverage at a low cost.
The key idea is to encourage individuals (users) to share their private residential Wi-Fi APs with each other, hence crowdsource the coverages of many private Wi-Fi APs \cite{communities, motivation}.
This can fully utilize the capacity of millions of Wi-Fi APs already installed, without requiring new installations by any single operator.
Meanwhile, by joining such a community network, each user can use not only his own AP (when staying at home\footnote{We use ``home'' to denote the location of the user's own Wi-Fi AP, which can correspond to residence, office, or even public areas (such as for those Wi-Fi provided by coffee shops).}), but also  other users' APs (when traveling outside).
Clearly, the success of such a crowdsourced network largely depends on the active participations and contributions of many Wi-Fi owners, and hence requires the careful design of a proper economic incentive mechanism.

One prominent commercial example of wireless community networks is FON \cite{FON}, which has more than 13 millions member Wi-Fi APs globally.\footnote{FON is especially popular in several European countries, such as UK, France, Belgium, and Netherlands, where FON provides good Wi-Fi coverage in almost all locations. It's also popular in South America and some East Asian countries (such as Japan and South Korea), where the FON provides good Wi-Fi coverage in several metropolitan areas.}
In FON, the operator incentivizes Wi-Fi AP owners (to share their private APs with others) by using two different incentive schemes, corresponding to two kinds of memberships: \emph{Linus} and  \emph{Bill}  \cite{LaFoneraUserManual}.
As a \emph{Linus}, a user does not receive any compensation when other users access his AP; meanwhile, he can use other FON members' APs free of charge.
As a \emph{Bill}, a user receives compensation when other users access his AP; meanwhile, he needs to pay for using other APs.
Moreover, if a user does not own a Wi-Fi AP, he can still access the FON network as an \emph{Alien}, who needs to pay for using any AP in the FON network.
The payments of Alien and Bill (for using other APs) are often \emph{time usage-based} (i.e., proportional to the Wi-Fi connection time) \cite{FONPass}.
Our study is motivated by the commercial successful example of FON. 

\subsection{Model and Contributions}\label{sec:contri}

\begin{figure}
\vspace{-5mm}
  \centering
   \includegraphics[width=0.45\textwidth]{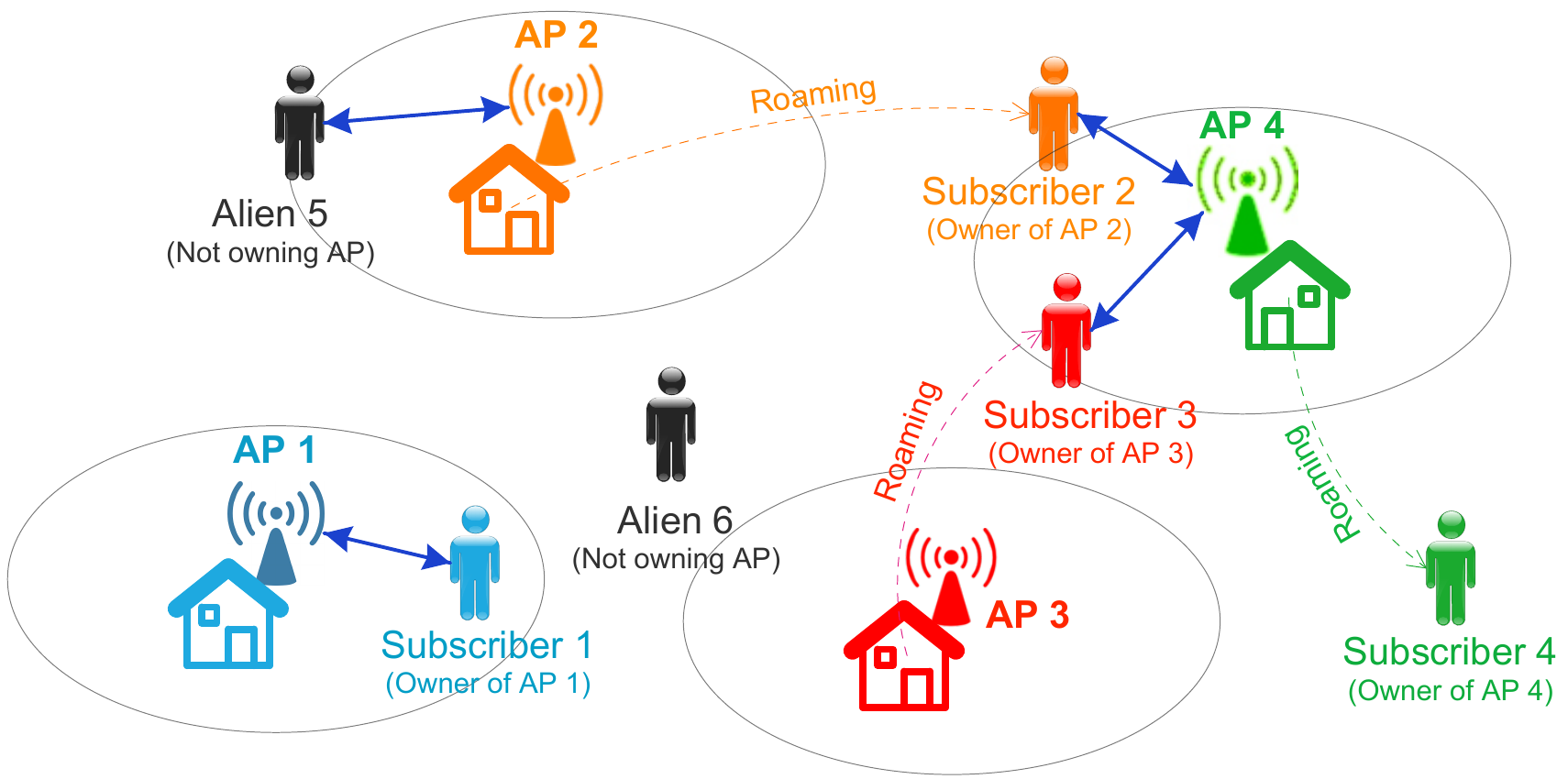}
  \caption{Wireless Community Network Model}\label{fig:model}
  \vspace{-5mm}
\end{figure}

\rev{
In this work, we consider a wireless community network launched by a network operator. 
The network consists of a set of users, including subscribers and Aliens.
}
Each subscriber is the owner of a private residential Wi-Fi AP associated with a specific home location, and shares his AP's Internet access with other users (subscribers and Aliens).
Similar as FON, we also assume that the network operator offers two types of memberships (i.e., \emph{Linus} and  \emph{Bill}) to its subscribers.
Each Alien does not own Wi-Fi AP, but can use subscribers' APs with a certain fee.
Users may travel (roam) outside his home location, and can use other subscribers' APs if needed.
Figure \ref{fig:model} illustrates such a wireless community network, where subscriber 1 (owner of AP 1) stays at home and connects to his own AP, subscribers 2 and 3 travel to subscriber 4's home location and connect to AP 4,
 Alien 5  travels to subscriber 2's home location and connects to AP 2,
 and subscriber 4 and Alien 6 roam at areas without Wi-Fi coverage and cannot connect to network.

The network operator and the users (subscribers and Aliens) interact in the following order.
First, the operator announces the pricing and incentive mechanism, i.e., the usage-based price charged to Bills and Aliens and the percentage of revenue shared with Bills.
Second, each subscriber chooses his membership type for a given \emph{time period} (e.g., six months), considering his mobility pattern within that time period as well as his demand and evaluation for network access during travel.
Third, if travelling to a particular
AP's location at a particular \emph{time slot} (e.g., five minutes), each user further decides his network access time on that AP during that time slot, taking the network congestion into consideration.
In this work, \emph{we will focus on the user decision problem}, given the pricing and incentive mechanism announced by the operator.

\begin{figure}
\vspace{-5mm}
  \centering
   \includegraphics[width=0.5\textwidth]{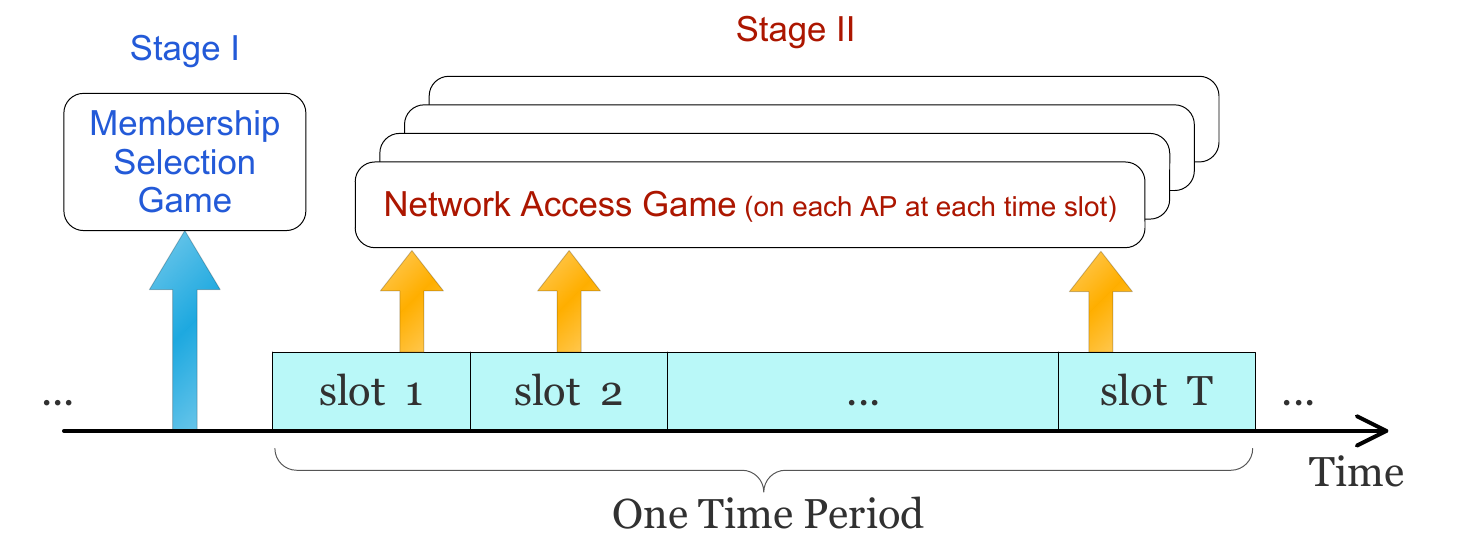}
  \caption{Two-Stage Game Model.
  At each time slot, there is a set of parallel network access games, each associated with an AP.}\label{fig:game}
    \vspace{-5mm}
\end{figure}

More specifically, we will study the users' joint membership selection and network access problem, and formulate the problem as a two-stage dynamic game.
In Stage I, subscribers choose their memberships (i.e.,  {Linus} or {Bill}) at the beginning of a time period, and all subscribers interact in a \emph{membership selection game}.
Each subscriber's membership choice will last for the whole time period.
In Stage II, at each time slot within the time period, each user decides his network access time on the AP at his current location (if he is not at home),
hence the users travelling to the same AP interact in a \emph{network access game}.
Since there are multiple APs in the network, we will have multiple concurrent network access games.
Figure \ref{fig:game} illustrates such a two-stage dynamic game model.
We analyze the Subgame Perfect Equilibrium (SPE) of this two-stage game systematically, and propose best response based algorithms to achieve the SPE.
We also provide numerical results to illustrate how the system parameters (e.g., users' mobility patterns and network access evaluations) affect the SPE.

The key contributions of this work are summarized as below.

\begin{itemize}
\item \emph{Novel Problem Formulation:} To the best of our knowledge, this is the first work that studies the users' joint membership selection and network access problem in a crowdsourced wireless community network.

\item \emph{Practical Relevance:}
Our model captures several key practical issues, such as the user mobility pattern,
network access evaluation,
demand response,
and network congestion effect,
which have not been fully considered before in the context of wireless community networks.

\item \emph{Equilibrium Analysis:}
We study the users' joint membership selection and network access problem from a game-theoretic perspective, and show how the user parameters affect their strategies under the game equilibrium: A user with a more popular home location, a smaller travel probability, or a smaller network access evaluation is more likely to choose to be a Bill.

\item \emph{Industry Insights:}
Our analysis can help the operator optimize the network pricing and incentive mechanism to achieve a maximum profit.

\end{itemize}

The rest of the paper is organized as below.
In Section \ref{sec:review}, we review the existing related literature.
In Section \ref{sec:model}, we present the system model.
In Sections \ref{sec:usa} and \ref{sec:membership}, we analyze the network access game in Stage II and membership selection game in Stage I, respectively.
We show simulation results and derive engineering insights in Section \ref{sec:simu}, and conclude in Section \ref{sec:conc}.  Due to space limit, most of the proofs are presented in the online technical report \cite{report}.

\section{Literature Review}\label{sec:review}

There are several closely related studies in wireless community networks, regarding incentive issues \cite{motivation}, the network expansion and interactions with traditional ISP \cite{Competition} \cite{Cooperation}, and the pricing mechanism design \cite{TwoPrice} \cite{Graph}.
Camponovo and Picco-Schwendener in \cite{motivation} concluded based on surveys that getting free Internet access from other members and revenue sharing are the two main incentives for users to join the FON network in Switzerland.
Manshaei \emph{et al.} in \cite{Competition} modeled a user's payoff as a function of the subscription fee and network coverage, and studied the evolution dynamics of wireless community networks.
Biczok \emph{et al.} in \cite{Cooperation} studied the competition and cooperation among users, wireless community network operator, and ISPs.
Authors in \cite{TwoPrice} \cite{Graph} focused on the pricing issues in wireless community networks.

In this work, we study both the membership selection and network access in a crowdsourced wireless community network.
Neither problem has been systematically studied in the existing literature.
Our model not only captures the Internet access sharing and revenue sharing as pointed out in \cite{motivation}, but also incorporates the impact of users mobility and the network congestion effect.
This makes our model and the derived insights more comprehensive and practically significant.

\section{System Model}\label{sec:model}

\subsection{The Network Model}

As illustarted in Figure \ref{fig:model}, we consider a crowdsourced wireless community network launched by a network operator, consisting of a set $\Ks =\{1,2,\ldots ,K\}$ of \emph{subscribers} (owning Wi-Fi APs) and a set $\Ka =\{K+1,K+2,...,K+K_{\textsc{a}}\}$ of \emph{Aliens} (not owning Wi-Fi APs).
We denote the set of all \emph{users} (including subscribers and Aliens) by $\Ku = \Ks \bigcup \Ka$.
A subscriber owns a private residential Wi-Fi AP and shares it with other users, while an Alien does not own Wi-Fi AP but can access subscribers' APs.
The network operator offers two memberships, i.e., Linus and Bill, to its subscribers, corresponding to different incentive schemes. Specifically,
\begin{itemize}
\item As a \emph{Linus}, a subscriber contributes his own AP without receiving compensation, and can use other APs free of charge;
\item As a \emph{Bill}, a subscriber needs to pay for using other APs, and can obtain a portion of the revenue collected at his AP by the operator.
\end{itemize}
An Alien has to pay for using any AP (as he does not contribute to the network).
For clarity, we summarize the above three user types in
Table \ref{table1}.

We consider a long time period (e.g., six months) consisting of $T$ time slots (e.g., five minutes per time slot).
Without loss of generality, we normalize the length of each time slot to be one.
Each subscriber makes the membership decision at the beginning of the time period, and cannot change such choice for the entire time period.
Users move randomly across time slots, and do not change their locations within a time slot.
Let $\eta_{i,j}$, $i\in\Ku, j\in\Ks$ denote the (stationary) probability that a (subscriber or Alien) user $i\in \Ku$ appears at the location of AP $j \in\Ks $ in any time slot, and $\eta_{i,0}$ denote the probability that user $i$ appears at a location that is not covered by any of the $K$ Wi-Fi APs.
We further define $\boldsymbol{\eta}_i=[\eta_{i,0},\eta_{i,1}, \ldots , \eta_{i,K}]$ as user $i$'s \emph{mobility pattern}. Obviously, $ \sum_{j=0}^K \eta_{i,j} =1 ,\ \forall i\in \Ku$.~~~~

To ensure a subscriber's Quality of Service (QoS) at his home location, each Wi-Fi AP splits the bandwidth into two separate channels (similar as the current practice of FON \cite{Channels}):
a \emph{Private Channel} for supporting its own communications,
and a \emph{Public Channel} for supporting roaming users' communications (from other subscribers and Aliens traveling to this location).
Hence, roaming users' communications will not interfere with a subscriber's own communication, and the network congestion only occurs among multiple users on the same public channel.

\subsection{The Operator and Users Interactions}

The operator and the users (both subscribers and Aliens) interact in the following order.

First, the network operator announces the pricing and incentive mechanism at the beginning of the time period, including
(i) the price  per unit connection time \cite{FON} paid by Aliens and Bills, denoted by $p \in (0, p_{\textsc{max}}]$,
and (ii) the percentage of revenue transferred to Bills, denoted by $\delta \in (0, 1)$.
In this paper, we will treat $(p, \delta)$ as fixed system parameters, and focus on studying the user behaviours.
This is because a full understanding of user behaviours is the first step towards the operator's optimal pricing and incentive mechanism design.
In Section \ref{sec:simu}, we will numerically illustrate how to properly  choose $(p,\delta)$ to optimize the operator's profit.

Second, given the operator's announcement $(p, \delta)$, each subscriber $i \in \Ks $ chooses his membership $x_i \in \{0, 1\}$ for the entire period of $T$ time slots, where $0 $ and $1$ correspond to ``Linus'' and ``Bill'', respectively.
The objective of each subscriber is to choose the best membership that maximizes his overall payoff during the period of $T$ time slots, considering users' mobility patterns as well as his demand and evaluation for network access (see Section \ref{sec:usa} for more details).

Third, given the operator's announcement $(p, \delta)$ and the subscribers' membership selections $\{x_i, i\in\Ks\}$, each user (subscriber or Alien) further decides the network usage in each time slot, i.e., the \emph{network access time} at the AP of his current location during  that time slot.
When staying at home, a subscriber uses his private channel exclusively, and his network access decision is independent of other users' decisions.
When accessing the Internet through another subscriber's AP, a user (subscriber or Alien) needs to compete for the public channel with other users at the same AP (except the owner of that AP),
hence his optimal network access time depends on other users' network access decisions.

In this work, \emph{we focus on the user decision problem}, i.e., the subscribers' membership selections and the users' network access decisions, given the pricing and incentive mechanism announced by the operator.

\begin{table}[t]
\vspace{-5mm}
\newcommand{\tabincell}[2]{\begin{tabular}{@{}#1@{}}#2\end{tabular}}
\centering
\caption{A Summary of Three User Types}
\begin{tabular}{|c|c|c|c|}
\hline
\textbf{User Type} & \tabincell{c}{\textbf{Pay for using other APs}} &  \tabincell{c}{\textbf{Paid by sharing his AP}}  \\
\hline
Linus & No & No \\
\hline
Bills & Yes & Yes \\
\hline
Aliens & Yes & Not Applicable  \\
\hline
\end{tabular}
\label{table1}
\vspace{-5mm}
\end{table}

\subsection{Game Formulation}

We formulate the joint membership selection and network access problem as a two-stage dynamic game, as illustrated in Figure \ref{fig:game}.
In Stage I, subscribers participate in a \emph{membership selection game} at the beginning of the whole time period, where each subscriber chooses his membership ({Linus} or {Bill}) for the whole time period.
In Stage II, at each time slot, users travelling to the same AP participate in a \emph{network access game}, where each user decides his network access time on that AP.
Namely, each AP is associated with a network access game at each time slot.

In what follows, we will analyze the two-stage game by backward induction, starting from Stage II.

\section{Stage II: Network Access Game on Each AP} \label{sec:usa}

We first study the network access game on a single AP at a single time slot in Stage II, given the subscribers' membership selections $\{x_i, i\in\Ks\}$ in Stage I and the operator's pricing and incentive mechanism $(p, \delta)$.
In this game, a user decides the network access time on the AP at his current location, aiming at maximizing his payoff in the current time slot.

\subsection{Network Access Game Formulation}

Without loss of generality, we consider the network access game on a particular AP $k$ at a particular time slot $t$. 
Recall that the length of a single time slot is normalized to be $1$. 

The \emph{players} are all users travelling to AP $k$ (except the owner of AP $k$) at time slot $t$, denoted by $\K(k, t) = \Ks(k,t) \bigcup \Ka(k,t)$, 
where $\Ks(k,t)$ and $\Ka(k,t)$ are the sets of subscribers and Aliens at this location and time, respectively.
For notational convenience, we will ignore the time index $t$ and write the player set as $\K(k) = \Ks(k) \bigcup \Ka(k)$ in the rest of this section, 
with the understanding that we already focus on a single time slot $t$.

The \emph{strategy} of each player $i \in \K(k)$ is to decide the network access time $\sigma_{i,k} \in [0,1] $ on AP $k$.
We denote the strategies of all players in $\K(k)$ except $i$ as $\boldsymbol{\sigma}_{-i,k} = \{\sigma_{j,k}, j\neq i, j\in\K(k) \}$.

The \emph{payoff} of player $i $ is a function of his own strategy $\sigma_{i,k}$ and other players' strategies $\boldsymbol{\sigma}_{-i,k} $, denoted by $v_{i,k}(\sigma_{i,k},\boldsymbol{\sigma}_{-i,k})$ (to be defined later).

The network access game on AP $k$ (at time slot $t$) and the corresponding Nash equilibrium are defined as follows.

\begin{game}[Network Access Game on AP $k$]\label{RCgame}
~
\begin{itemize}
\item Players: the set $ \K(k) $ of users;
\item Strategies: $\sigma_{i,k} \in [0,1]$, $\forall  i \in \K(k) $;
\item Payoffs: $v_{i,k}(\sigma_{i,k},\boldsymbol{\sigma}_{-i,k})$, $\forall  i \in \K(k) $.
\end{itemize}
\end{game}

\begin{definition}\label{NEII}
A Nash equilibrium of the Network Access Game on AP $k$ (at time slot $t$) is a profile $\boldsymbol{\sigma}_k^\ast=\{\sigma_{i,k}, \forall i\in \K(k)\}$ such that for each user $ i \in \K(k) $,
\begin{equation*}
v_{i,k}(\sigma_{i,k}^\ast,\boldsymbol{\sigma}_{-i,k}^\ast) \geq  v_{i,k}(\sigma_{i,k},\boldsymbol{\sigma}_{-i,k}^\ast),\quad \forall \sigma_{i,k} \in [0,1].
\end{equation*}
\end{definition}

Note that the Nash equilibrium $\boldsymbol{\sigma}_k^\ast$ depends on the player set $\K(k)$, hence can be written as $\boldsymbol{\sigma}_k^\ast (\K(k))$.

\subsection{Utility and Payoff Definition}

Before analyzing the Nash equilibrium, we first define users' utility and payoff functions.

\subsubsection{\textbf{Utility}}
The \emph{utility} captures a user's satisfaction for accessing the Internet for a certain amount of time.
Due to the principle of diminishing marginal returns \cite{logutility, logutility2}, the utility function is often increasing and concave.
As a concrete example, we define the utility of user $i\in \K(k)$ on AP $k$ as
\begin{equation}
u_i(\sigma_{i,k},\boldsymbol{\sigma}_{-i,k}) = \rho_i \log(1+\bar{r}_{i,k}(\boldsymbol{\sigma}_{-i,k}) \cdot  \sigma_{i,k}). \label{uti}
\end{equation}
Here $\rho_i$ is the \emph{network access evaluation} of user $i$, characterizing user $i$'s valuation of data consumption.
Furthermore, $\bar{r}_{i,k}(\boldsymbol{\sigma}_{-i,k})$ is the expected data rate that user $i$ can achieve on AP $k$, which is a decreasing function of other users' network access vector $\boldsymbol{\sigma}_{-i,k}$.
Intuitively, if more users access AP $k$'s public channel simultaneously,  user $i$'s achieved data rate will decrease due to the increased congestion.
Obviously, $\bar{r}_{i,k}(\boldsymbol{\sigma}_{-i,k}) \cdot  \sigma_{i,k} $ denotes the total expected amount of data that user $i$ consumes on AP $k$ (at time slot $t$).

Next, we derive the user $i$'s expected data rate  $\bar{r}_{i,k}(\boldsymbol{\sigma}_{-i,k})$.
Let $\bar{R}(n)$ denote the average data rate of a Wi-Fi user when $n$ users connect to the same Wi-Fi AP simultaneously.
Let $P_{i,k} (n )$ denote the probability that $n $ other users (except $i$) connect to AP $k$.
Then,  user $i$'s expected data rate $\bar{r}_{i,k}(\boldsymbol{\sigma}_{-i,k})$ at AP $k$  can be calculated as follows:
\begin{equation}
\bar{r}_{i,k}(\boldsymbol{\sigma}_{-i,k})=\sum_{n=0}^{|\K(k)|-1} P_{i,k} (n) \cdot \bar{R}(n+1). \label{DataRate}
\end{equation}

\begin{figure}
  \vspace{-5mm}
  \centering
   \includegraphics[width=0.28\textwidth]{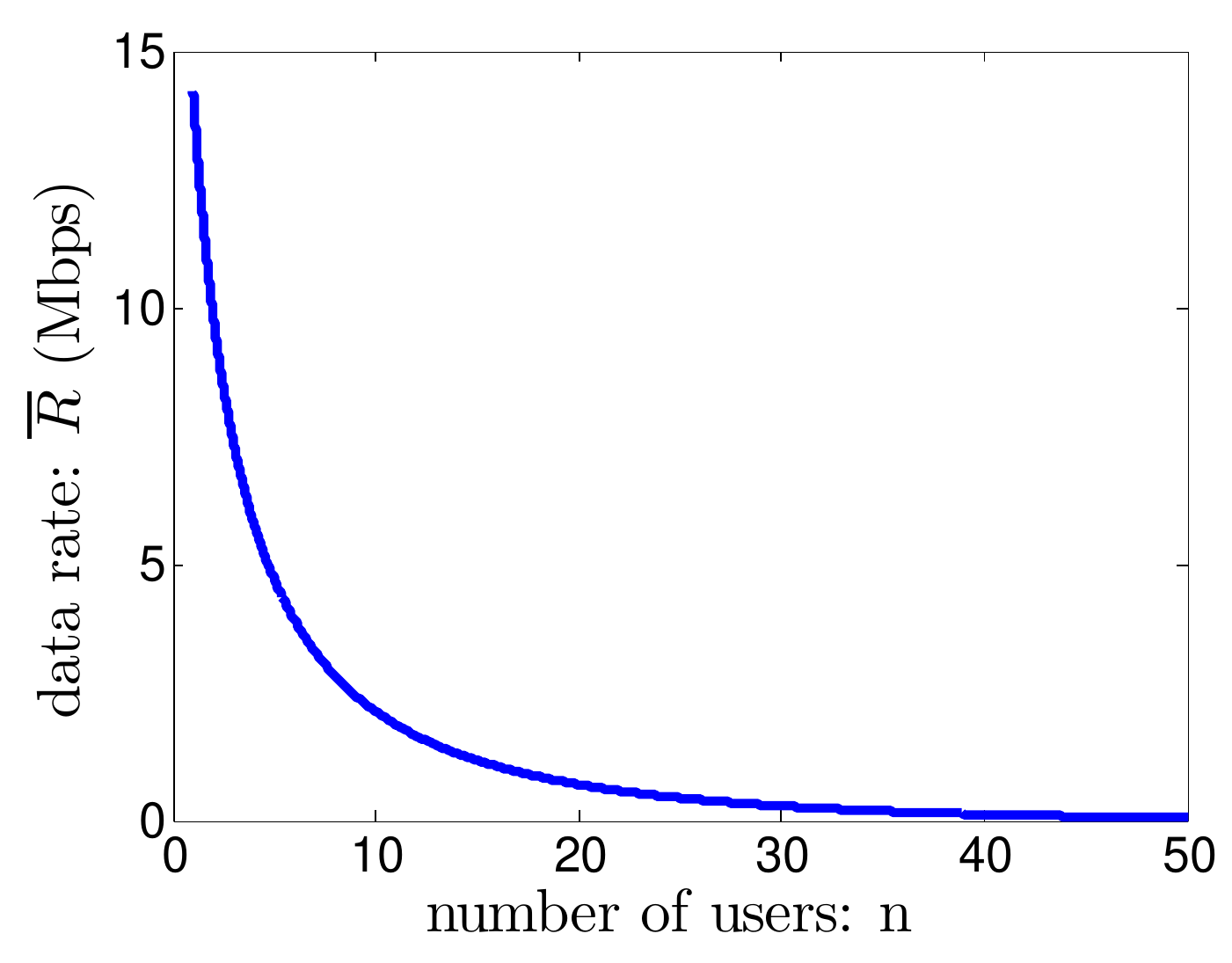}
  \caption{Average Data Rate per User \cite{20Q}}\label{fig:DataRate}
\vspace{-5mm}
\end{figure}

According to IEEE $802.11$g standard \cite{20Q}, we have:
\begin{equation*}
\bar{R}(n)=
\frac{\tau \bar{\tau}^{n-1}L}{\bar{\tau}^nT_b+[(1-\bar{\tau}^n)-n\tau\bar{\tau}^{n-1}]T_c+n\tau\bar{\tau}^{n-1}T_s} , 
\end{equation*}
where $\tau$ is the average successful probability of contention (and $\bar{\tau} = 1-\tau$), $L$ is the average payload length, $T_b$ is the length of a backoff slot, $T_c$ is the length of a collision slot, and $T_s$ is the length of a successful slot.
Figure \ref{fig:DataRate} illustrates an example of $\bar{R}(\cdot)$ under IEEE $802.11$g standard (reproduced from \cite{20Q}, with parameters $\tau=0.0765, L=8192, T_b=28\mu s$, and $T_c=T_s=85.7+L/54 \mu s$).
The decreasing data rate per user is due to both the reduced resource per user and the waste of resources caused by congestion among users.

For simplicity, we further assume that if a user $i$ decides to connect to the channel with a total time of $\sigma_{i,k}\in[0,1]$, he will spread this access time \emph{randomly and uniformly} across the entire time slot.
Recall that the length of a time slot is normalized to $1$.
Hence, the probability that user $i$ connects to AP $k$
\emph{in an infinitely small time interval within the time slot} is $\sigma_{i,k}$.
Therefore, the probabilities $P_{i,k} (n), n=0,1,...,|\K(k)|-1$, follow the binomial distribution (with a total of $|\K(k)|$ trials and a success probability $\sigma_{j,k}$ for each trial $ j \in \K(k)/\{i\}$).
Formally, we have:
\vspace{-8pt}
\begin{equation*}
P_{i,k} (n ) = \sum_{\K_{n } \in \mathbf{K}_{n }(k)}
\left(
\prod_{j \in \K_{n } } \sigma_{j,k} \cdot  \prod_{j \in \K(k)/\{i\}/\K_{n } } ( 1- \sigma_{j,k})
\right),
\end{equation*}
where $\K_{n } $ denotes an arbitrary subset of $\K(k)$ with $n$ users (except $i$), and $\mathbf{K}_{n }(k)$ denotes the set of all  possible $\K_{n } $.
Obviously, $\prod_{j \in \K_{n } } \sigma_{j,k} $ denotes the probability that all users in $\K_{n }  $ are connecting to AP $k$, and $\prod_{j \in \K(k)/\{i\}/\K_{n } } ( 1- \sigma_{j,k})$ denotes the probability that all other users (except user $i$ and those in $\K_{n } $)  are \emph{not} connecting to AP $k$.

\subsubsection{\textbf{Payoff}}

The \emph{payoff} of each user $i\in \K(k)$ in the network access game on AP $k$ (at time slot $t$) is defined as the difference between the utility and the payment (charged to Bills and Aliens).

Specifically, if user $i$ is a Linus (i.e., $i\in \Ks (k)$ and $x_i = 0 $),
he does not need to pay for his network usage at AP $k$.
Hence, the payoff of a Linus-type user $i$ on AP $k$, denoted by $v_{i,k}^\L$, is the same as his utility defined in \eqref{uti}, i.e.,
\begin{equation}
v_{i,k}^\L (\sigma_{i,k},\boldsymbol{\sigma}_{-i,k}) = u_i(\sigma_{i,k},\boldsymbol{\sigma}_{-i,k}). \label{payoffLinus}
\end{equation}
If user $i$ is a Bill (i.e., $i\in \Ks(k)$ and $x_i = 1$) or Alien (i.e., $i\in \Ka (k) $),
he needs to pay for his network usage on AP $k$, and the payment is proportional to his network access time $\sigma_{i,k}$.
Hence, the {payoff} of a Bill-type or Alien user $i$, denoted by $v_{i,k}^\B$, is the difference between utility and payment, i.e.,
\begin{equation}
v_{i,k}^\B(\sigma_{i,k},\boldsymbol{\sigma}_{-i,k}) = u_i(\sigma_{i,k},\boldsymbol{\sigma}_{-i,k})- p \sigma_{i,k} . \label{payoffBill}
\end{equation}

Based on the above, we can summarize the payoff of user $i\in \K(k)$ in the Network Access Game (on AP $k$) as follows:
$$
v_{i,k} (\sigma_{i,k},\boldsymbol{\sigma}_{-i,k}) = ~~~~~~~~~~~~~~~~~~~~~~~~~~~~~~~~~~~~~~
$$
\begin{equation}\label{eq:payoff}
\left\{
\begin{aligned}
v_{i,k}^\L (\sigma_{i,k},\boldsymbol{\sigma}_{-i,k}), &\quad \mbox{ if } i \in \Ks(k) \mbox{ and } x_i = 0;
\\
v_{i,k}^\B (\sigma_{i,k},\boldsymbol{\sigma}_{-i,k}), & \quad \mbox{ if } i \in \Ks(k) \mbox{ and } x_i = 1;
\\
v_{i,k}^\B (\sigma_{i,k},\boldsymbol{\sigma}_{-i,k}), & \quad \mbox{ if } i \in \Ka(k).
\end{aligned}
\right.
\end{equation}

\subsection{Nash Equilibrium Analysis}

Now we study the Nash equilibrium of the above Network Access Game (on AP $k$) systematically.\footnote{Due to space limit, we put most of the detailed proofs in the online technical report \cite{report}.}

Given all other users' strategies, a user's \emph{best response} is the strategy that maximizes his payoff. The Nash equilibrium is a strategy profile where each user's strategy is the best response to other users' strategies.

\begin{lemma}\label{lemmaBRLinus}
If user $i$ is a Linus, his best response in the Network Access Game on AP $k$ is
\begin{equation}
\sigma_{i,k}^\ast = 1,
\label{BRLinus}
\end{equation}
regardless of other users' strategies.
\end{lemma}


\begin{lemma}\label{lemmaBRBill}
If user $i$ is a Bill or an Alien, his best response in the Network Access Game on AP $k$ is
\begin{equation}
\sigma_{i,k}^\ast = \min\left\{ 1,\max\left\{ \frac{\rho_i}{p}-\frac{1}{\bar{r}_{i,k}(\boldsymbol{\sigma}_{-i,k})},0 \right\} \right\},  \label{BRBill}
\end{equation}
which is a function of other users' strategies $\boldsymbol{\sigma}_{-i,k}$.
\end{lemma}


We next give the existence of the Nash equilibrium in the Network Access Game.

\begin{theorem}\label{ExiNEI}
There exists at least one Nash equilibrium in the Network Access Game on AP $k$.
\end{theorem}


Now we discuss the uniqueness of the Nash equilibrium in the Network Access Game on AP $k$.

\begin{proposition}\label{UniNEI}
In a Network Access Game with two players, the Nash equilibrium is unique if
$ \frac{\bar{R}(1)-\bar{R}(2)}{(\bar{R}(2))^2} < 1 .$
\end{proposition}

Note that the condition in Proposition \ref{UniNEI} is always satisfied for practical WiFi systems \cite{20Q}.
For the cases with more than two players, however, the uniqueness of the Nash equilibrium depends on the system parameters in a more complicated fashion.
Please refer to our technical report \cite{report} for more detailed discussions.


We further propose a best response update algorithm, which is guaranteed to linearly converge to the Nash equilibrium under the same condition for the uniqueness of the Nash equilibrium. For details, see \cite{report}.

\section{Stage I: Membership Selection Game} \label{sec:membership}

Now we study the subscribers' membership selection game in Stage I, given the operator's pricing and incentive mechanism $(p, \delta)$.
In this stage, each subscriber $i\in\Ks$ decides his membership type $x_i \in \{0, 1\}$ (i.e., Linus or Bill) at the beginning of the period, aiming at maximizing the overall payoff that he can achieve in all $T$ time slots.
Note that an Alien $i \in \Ka$ cannot choose his type, as he has no Wi-Fi AP and does not contribute to the network.

\subsection{Membership Selection Game Formulation}

In the Membership Selection Game, \emph{players} are all subscribers in the set $\Ks$. The \emph{strategy} of each player $i\in \Ks$ is to decide his membership $x_i \in \{0, 1\}$, with $x_i = 0$ and $1$ denoting Linus and Bill, respectively.
Such a membership choice will last for the whole time period.
We denote the strategies of all players except $i$ by $\boldsymbol{x}_{-i} = \{x_j, j\neq i, j \in \Ks\}$.
The \emph{overall payoff}
of a player $i$ is sum of the \emph{total expected payoff} on all APs that he may travel to and the \emph{total expected revenue} that he may collect at his own AP (if choosing to be a Bill) during $T$  slots.
Obviously, it is a function of his own strategy $x_i$ and other players' strategies $\boldsymbol{x}_{-i}$, denoted by $V_i(x_i,\boldsymbol{x}_{-i})$  (to be defined later).

Formally, the Membership Selection Game and the corresponding Nash equilibrium can be defined as follows.

\begin{game}[Membership Selection Game]\label{MSgame}
$\mbox{ }$
\begin{itemize}
\item Players: the set $\Ks$ of subscribers.
\item Strategies: $x_i\in \{0,1\}$, $\forall i \in \Ks$.
\item Payoffs: $V_i(x_i,\boldsymbol{x}_{-i})$, $\forall i \in \Ks$.
\end{itemize}
\end{game}

\begin{definition}\label{NEI}
A Nash equilibrium of the Membership Selection Game is a profile $\boldsymbol{x}^\ast=\{x_i^\ast,i \in \Ks\}$ such that for each subscriber $i \in \Ks$,
$$
V_i(x_i^\ast,\boldsymbol{x}_{-i}^\ast) \geq  V_i(x_i,\boldsymbol{x}_{-i}^\ast),\quad \forall x_i \in\{0, 1\}.
$$
\end{definition}

We note that the Nash equilibria in Stage II (Definition \ref{NEII}) and Stage I (Definition \ref{NEI}) together form a Subgame Perfect Equilibrium (SPE) of the whole game.

\subsection{Payoff Definition}

Before analyzing the Nash equilibrium, we first explicitly calculate each subscriber's overall payoff during the whole period, which includes (i) the total expected payoff on all APs that he may travel to, and (ii) the potential revenue that he may collect on his own AP (if choosing to be a Bill).

We first calculate the total expected payoff of each subscriber (on all APs that he may travel to), which depends on his mobility pattern.
Recall that the mobility of a subscriber $i$ is characterized by the probabilities of travelling to different APs, i.e., $\boldsymbol{\eta}_i=[\eta_{i,0},\eta_{i,1},\ldots , \eta_{i,K}]$, where $\eta_{i,k} $ is the probability of subscriber $i$ travelling to AP $k$, and $\eta_{i,0}$ is the probability of subscriber $i$ travelling to an area that is not covered by any of the $K$ Wi-Fi APs.
We calculate subscriber $i$'s expected payoffs (per time slot) when staying at home and when roaming outside, respectively.

\subsubsection{Stay at home (with probability $\eta_{i,i}$)}

When staying at home, subscriber $i$ communicates over the private channel of AP $i$ and does not interfere with other users.
Hence his expected payoff, denoted by $V_{i,i}(x_i,\boldsymbol{x}_{-i})$, is
\begin{equation*}
V_{i,i}(x_i,\boldsymbol{x}_{-i}) =  \rho_i \cdot \log(1+\bar{r}_{i} \cdot 1),  \label{VkLinus-x}
\end{equation*}
where constant $\bar{r}_{i} $ corresponds to the average data rate achieved at his private channel.
The product term $\bar{r}_{i} \cdot 1$ implies that user $i$ will access the Internet during the entire time slot.

\subsubsection{Travel to AP $k \neq i$ (with probability $\eta_{i,k}$)}

When travelling to another AP $k\neq i$, subscriber $i$ needs to compete over the public channel with other users (except $k$) travelling to AP $k$ at the same time (in the Network Access Game).

Suppose that a set $\M(k)$ of other users (except $i$ and $k$) are travelling to AP $k$ at the same time.
That is, the game player set in the Network Access Game on AP $k$ is $\K(k) = \M(k) \bigcup \{i\}$.
For more clarity, let us rewrite the equilibrium payoff of subscriber $i$ on AP $k$, i.e.,  $v_{i,k} (\sigma_{i,k},\boldsymbol{\sigma}_{-i,k})$ defined in \eqref{eq:payoff}, as $v_{i,k} (\sigma_{i,k},\boldsymbol{\sigma}_{-i,k}| \M (k))$, \emph{when competing with a set $\M(k)$ of other users (in the Network Access Game on AP $k$)}.
Hence, the expected payoff of subscriber $i$ on AP $k$ is
\begin{equation*}
V_{i,k}(x_i,\boldsymbol{x}_{-i}) =  \sum_{\M(k) \in \mathbf{K}_{-\{i,k\}} } \phi (\M(k)) v_{i,k} (\sigma_{i,k}^\ast,\boldsymbol{\sigma}_{-i,k}^\ast| \M(k)),
\end{equation*}
where
$\phi (\M(k))$ is the probability that a set $\M(k)$ of users are travelling to AP $k$,
$(\sigma_{i,k}^\ast, \boldsymbol{\sigma}_{-i,k}^\ast )$ is the corresponding equilibrium of the Network Access Game,
and $\mathbf{K} _{-\{i,k\}} $ is the power set of {$\Ku/\{i,k\}$}, i.e., the set of all subsets of $\Ku/\{i,k\}$.
The probability $\phi (\M(k))$ is given by\footnote{\rev{In this work, we study the complete information scenario where users' mobility patterns are public information.}}
\begin{equation*}
\phi (\M(k)) = \prod_{j \in \M(k)} \eta_{j,k} \cdot \prod_{j \in \Ku / \{i,k\}/\M(k) } (1 - \eta_{j,k}),
\end{equation*}
where $\prod_{j \in \M(k)} \eta_{j,k} $ denotes the probability that all users in $\M(k) $ are travelling to AP $k$, and $\prod_{j \in \Ku/\{i,k\}/\M(k)} (1 - \eta_{j,k}) $ denotes the probability that all other users (except users $i$, $k$, and those in $\M(k)$)  are \emph{not} travelling to AP $k$.

\subsubsection{Travel outside the network coverage (with probability $\eta_{i,0}$)}

When travelling to an area that is not covered by any of the $K$ Wi-Fi APs, the expected payoff of subscriber $i$, denoted by $V_{i,0}(x_i,\boldsymbol{x}_{-i}) $, is\footnote{If a user can access the Internet through other means, we can normalize the corresponding constant payoff to be zero without affecting the analysis.}
\begin{equation*}
V_{i,0}(x_i,\boldsymbol{x}_{-i}) = 0.
\end{equation*}

Based on the above, the total expected payoff of subscriber $i$ (on all APs that he may travel to during the whole period of $T$ time slots) is
\begin{equation}\label{eq:totalpayoff-1}
V_i^\dag (x_i,\boldsymbol{x}_{-i}) = T \cdot \sum_{k=0}^K
\eta_{i, k} \cdot
V_{i,k}(x_i,\boldsymbol{x}_{-i}) .
\end{equation}

Next, we calculate the total expected potential revenue that each subscriber $i$ may collect on his own AP.
Specifically, if choosing to be a Linus, subscriber $i$ obtains a zero revenue from his AP.\footnote{The operator still charges Bills and Aliens for using a Linus' AP.}
If choosing to be a Bill, subscriber $i$ obtains a fixed portion $\delta$ of the revenue collected at his AP.

Suppose that a set $\K(i)$ of other users  (except $i$) are travelling to AP $i$. That is, the player set in the Network Access Game on AP $i$ is $\K(i)$.
Then, the Nash equilibrium in the Network Access Game on AP $i$ can be written as $\{ \sigma_{j,i}^\ast (\K(i)), \forall j\in \K(i)\}$.
Recall that the revenue collected at each AP is the total payment of all Aliens and Bills (except the owner of that AP) accessing that AP.
Hence, the total revenue collected at AP $i$ is
\begin{equation*}
\begin{aligned}
\Pi_i (\boldsymbol{x}_{-i}, \K(i)) = &   \sum_{j \in \K(i) \bigcap \Ka} p \cdot \sigma_{j,i}^\ast (\K(i))
\\
 &  + \sum_{j \in \K(i) \bigcap \Ks} x_j \cdot p \cdot  \sigma_{j,i}^\ast (\K(i)) ,
\end{aligned}
\end{equation*}
where the first term is the payment of Aliens, and the second term is the payment of Bills.
Hence, the total expected payment of Bills and Aliens at AP $i$ is
\begin{equation*}
\bar{\Pi}_i (\boldsymbol{x}_{-i}) = \sum_{\K(i) \in \mathbf{K}_{-i} } \psi (\K(i)) \cdot \Pi_i (\boldsymbol{x}_{-i}, \K(i)) ,
\end{equation*}
where $\psi (\K(i)) $ is the probability that a set $\K(i)$ of users are travelling to AP $i$, and $\mathbf{K} _{-i} $ is the power set of $\Ku/\{i\}$.
The probability $\psi (\K(i))$ is given by
\begin{equation*}
\psi (\K(i)) = \prod_{j \in \K(i)} \eta_{j,i} \cdot \prod_{j \in \Ku / \{i\}/\K(i)} (1 - \eta_{j,i}).
\end{equation*}

Based on the above, the total expected revenue that a Bill subscriber $i$ can achieve at his own AP (during the whole time period of $T$ time slots) is
\begin{equation}\label{eq:totalpayoff-2}
V_i^\ddag (x_i,\boldsymbol{x}_{-i}) = x_i \cdot T \cdot \delta \cdot \bar{\Pi}_i  (\boldsymbol{x}_{-i}) .
\end{equation}

Combining the total expected payoff in \eqref{eq:totalpayoff-1} and the total expected revenue in \eqref{eq:totalpayoff-2}, we obtain the {overall payoff} of each subscriber in the Membership Selection Game as follows
\begin{equation}\label{eq:totalpayoff}
\begin{aligned}
 & V_i (x_i,\boldsymbol{x}_{-i})  = V_i^\ddag (x_i,\boldsymbol{x}_{-i}) +
V_i^\dag (x_i,\boldsymbol{x}_{-i})
\\
& = T \cdot \left(
x_i \cdot \delta \cdot \bar{\Pi}_i  (\boldsymbol{x}_{-i})
+
\sum_{k=0}^K
\eta_{i, k} \cdot
V_{i,k}(x_i,\boldsymbol{x}_{-i})  \right).
\end{aligned}
\end{equation}

\subsection{Nash Equilibrium Analysis}

A subscriber $i$ will make the membership decision to maximize the overall payoff defined in \eqref{eq:totalpayoff}.
Specifically, he will choose to be a Linus if $V_i(0,\boldsymbol{x}_{-i}) > V_i(1,\boldsymbol{x}_{-i})$, and choose to be a Bill otherwise.
For notational convenience, we denote $f_i(\boldsymbol{x}_{-i})$ as the gap between  $V_i(1,\boldsymbol{x}_{-i}) $ and $ V_i(0,\boldsymbol{x}_{-i})$:
\begin{equation}
f_i(\boldsymbol{x}_{-i}) = V_i(1,\boldsymbol{x}_{-i}) - V_i(0,\boldsymbol{x}_{-i}). \label{f}
\end{equation}
Hence, subscriber $i$ will choose to be a Linus if $f_i(\boldsymbol{x}_{-i}) < 0$, and choose to be a Bill if $f_i(\boldsymbol{x}_{-i}) \geq  0$.
Mathematically, this is equivalent to choosing $x_i$ from $ \{0 , 1\}$, such that the following condition holds:
$
(2x_i-1)\cdot f_i(\boldsymbol{x}_{-i}) \geq 0.
$

Next, we study the Nash equilibrium of the Membership Selection Game.

\begin{lemma}\label{lemmaMS}
A membership profile $\boldsymbol{x}^\ast$ is an Nash equilibrium of the Membership Selection Game, if and only if
$$
(2x_i^\ast-1)\cdot f_i(\boldsymbol{x}_{-i}^\ast) \geq 0, \quad  \forall i\in \mathcal{K}.
$$
\end{lemma}


\begin{proposition}\label{lemmaETA}
For each subscriber $i$, if
$$
\eta_{i,i} > \underline{\eta}_{i} \triangleq
1 - \frac{\delta \cdot \bar{\Pi}_i  (\boldsymbol{x}^\ast_{-i})}{
\sum_{k\in \Ks/\{i\}} \left(V_{i,k}(0,\boldsymbol{x}^\ast_{-i}) - V_{i,k}(1,\boldsymbol{x}^\ast_{-i})\right)} ,
$$
choosing Bill (i.e., $x_i=1$) is his best response.
\end{proposition}


Intuitively, a subscriber with a large probability of staying at home will choose to be a Bill, as his network usage on other APs is small, hence the benefit of obtaining revenue at his own AP outweighs the payment at other APs.

Unfortunately, the above Membership Selection Game may not always have an Nash equilibrium defined in Definition \ref{NEI}, which is a \emph{pure strategy} equilibrium where each subscriber chooses either to be a Bill or a Linus.
To illustrate this, we provide a simple example with 3 APs in our technical report \cite{report}.
Hence, in what follows, we will further look at the case of \emph{mixed-strategy} Nash equilibrium, where each subscriber chooses membership with probability.

\subsection{Mixed-Strategy Nash Equilibrium}\label{subsec:mexed}

For each subscriber $i$, his \emph{mixed strategy} can be characterized as the probability $\alpha_i \in [0,1]$ of choosing to be a Bill (hence the probability of choosing to be a Linus is  $1 - \alpha_i$).
Obviously, the pure strategy $x_i$ is a special case of the mixed strategy when $\alpha_i$ equals 1 or 0.
For writing convenience, we denote the mixed strategy profile of all subscribers except $i$ as
$$
\boldsymbol{\alpha}_{-i}=\{\alpha_j,j\neq i, j \in \Ks\}.
$$
Then, the expected payoff of subscriber $i$ can be defined as
\begin{equation}
\omega_i(\alpha_i,\boldsymbol{\alpha}_{-i})=\alpha_i \cdot \widetilde{V}_i(1,\boldsymbol{\alpha}_{-i})  + (1-\alpha_i) \cdot \widetilde{V}_i(0,\boldsymbol{\alpha}_{-i}) ,  \label{MixPayoff}
\end{equation}
where $\widetilde{V}_i(1,\boldsymbol{\alpha}_{-i})$ and $\widetilde{V}_i(0,\boldsymbol{\alpha}_{-i})$ are expected payoffs of subscriber $i$ when choosing to be a Bill and a Linus, respectively.
Note that $\widetilde{V}_i(1,\boldsymbol{\alpha}_{-i})$ and $\widetilde{V}_i(0,\boldsymbol{\alpha}_{-i})$ are the expected values on all possible membership selections of other users.
Specifically, there are $K-1$ other subscribers, hence $2^{K-1}$ possible membership selection combination of those subscribers, forming a set $\mathcal{X}_{-i}$.
Notice that each subscriber $j$ chooses $x_j = 1$ and $0$ with probabilities $\alpha_j$ and  $1 - \alpha_j$, respectively. Then, the probability that a particular $\boldsymbol{x}_{-i}\in\mathcal{X}_{-i} $ is realized is
$$
\psi (\boldsymbol{x}_{-i}) = \prod_{j \in \Ks/\{i\}}
\big( \alpha_j \cdot x_j + (1 - \alpha_j) \cdot (1 - x_j) \big) .
$$
Then, $\widetilde{V}_i(1,\boldsymbol{\alpha}_{-i})$ and $\widetilde{V}_i(0,\boldsymbol{\alpha}_{-i})$ can be calculated by
\begin{align}
& \widetilde{V}_i(x_i,\boldsymbol{\alpha}_{-i})
=
\sum_{\boldsymbol{x}_{-i} \in\mathcal{X}_{-i}}
\psi (\boldsymbol{x}_{-i})  \cdot  V_i(x_i,\boldsymbol{x}_{-i}) ,~~ x_i\in\{0,1\}, \notag
\end{align}
where $V_i(x_i,\boldsymbol{x}_{-i})$ is the overall payoff of subscriber $i$ under the pure strategy profile defined in \eqref{eq:totalpayoff}.

\begin{definition}\label{NEImixed}
A mixed-strategy Nash equilibrium of the Membership Selection Game is a probability profile $\boldsymbol{\alpha}^\ast$ such that for each subscriber $i \in \Ks$, we have:
$$ \omega_i(\alpha_i^\ast,\boldsymbol{\alpha}_{-i}^\ast) \geq  \omega_i(\alpha_i,\boldsymbol{\alpha}_{-i}^\ast), \quad \forall \alpha_i\in[0,1]. $$
\end{definition}

We first show the existence of the mixed-strategy Nash equilibrium in the Membership Selection Game.

\begin{theorem}\label{lemmaExiMixedNE}
In the Membership Selection Game, there exists at least one mixed-strategy Nash equilibrium.
\end{theorem}


Similarly, to compute the Nash equilibrium effectively,
We design a \emph{smoothed} best response updated algorithm, where each player updates his mixed strategy in a smoothed best response manner according to the other players' mixed strategies in the previous iteration.
Using the result in \cite{Game-Learning}, we can show that such a smoothed best response with some learning rules (as in fictitious play) converges to the mixed strategy Nash equilibria.
Details of the algorithm is shown in Appendix \cite{report}.

\begin{figure*}[t]
\vspace{-5mm}
\centering
\begin{minipage}[t]{0.3 \linewidth}
\centering
\includegraphics[height=1.6 in]{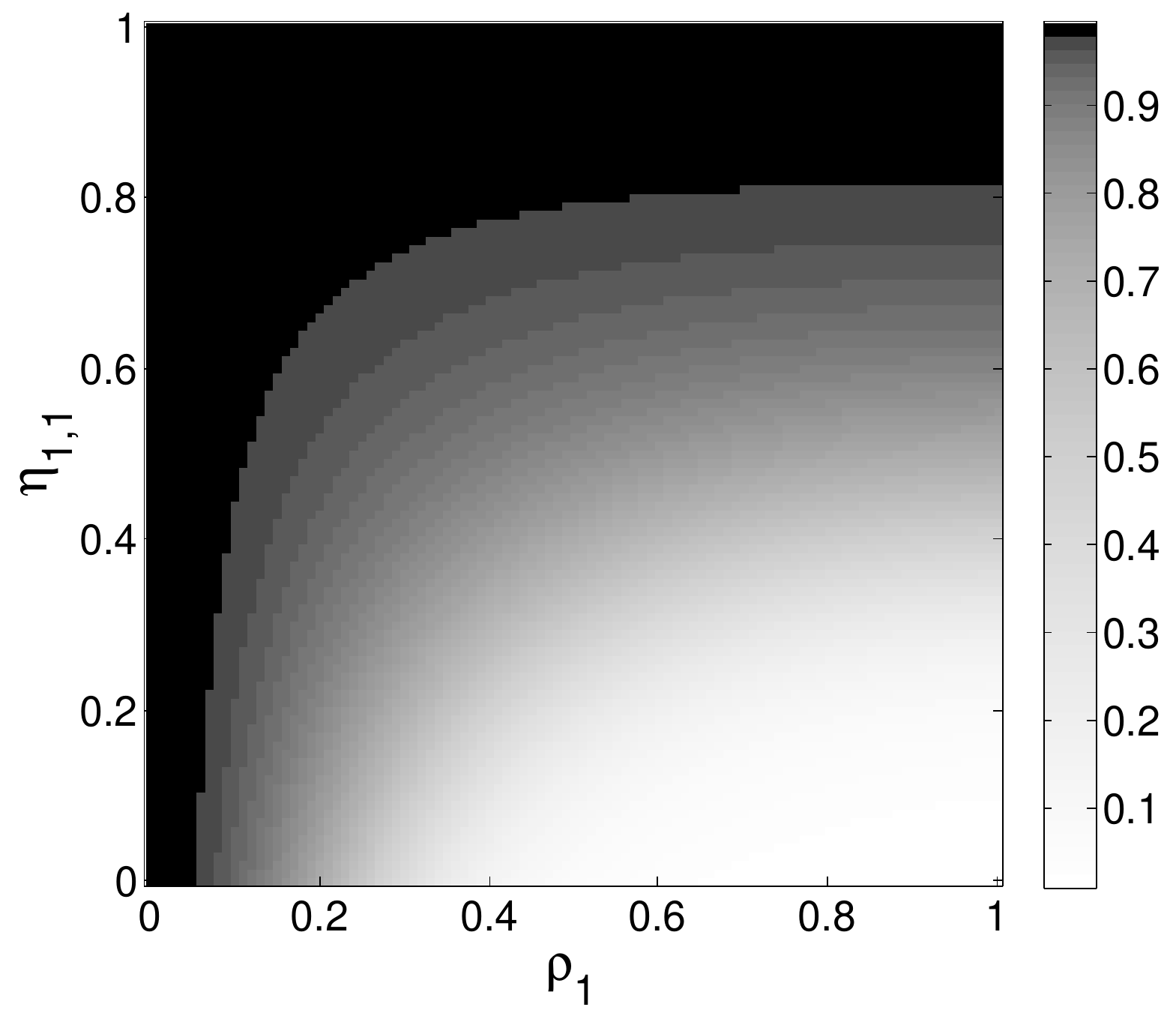}
  \caption{Membership Decision under Different Parameters}\label{fig:Parameter}
\end{minipage}
\begin{minipage}[t]{0.03 \linewidth}
~
\end{minipage}
\begin{minipage}[t]{0.3 \linewidth}
\centering
\includegraphics[height=1.6 in]{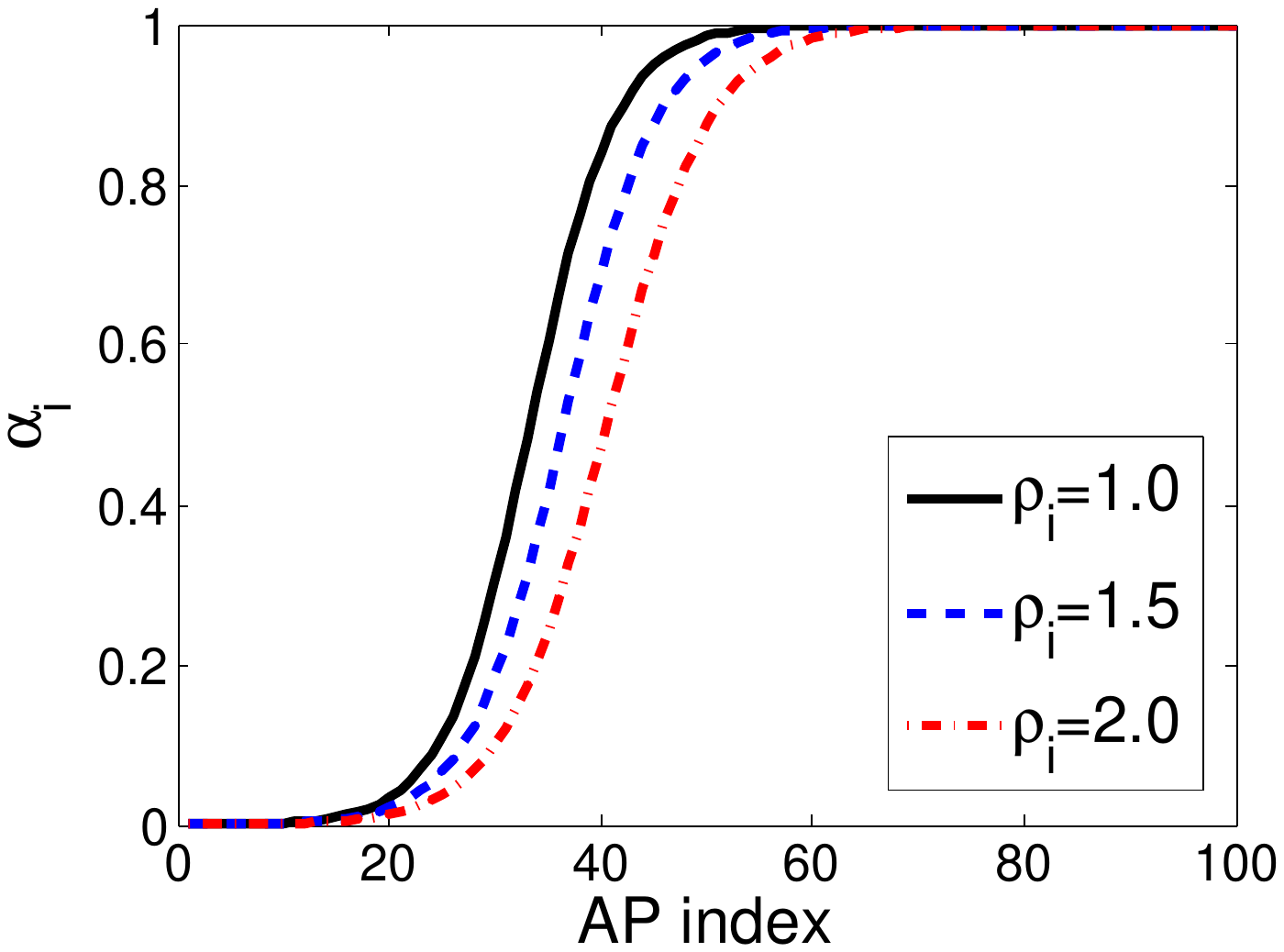}
  \caption{Membership Decisions of All Subscribers}\label{fig:LP}
\end{minipage}
\begin{minipage}[t]{0.03 \linewidth}
~
\end{minipage}
\begin{minipage}[t]{0.3 \linewidth}
\centering
\includegraphics[height=1.6 in]{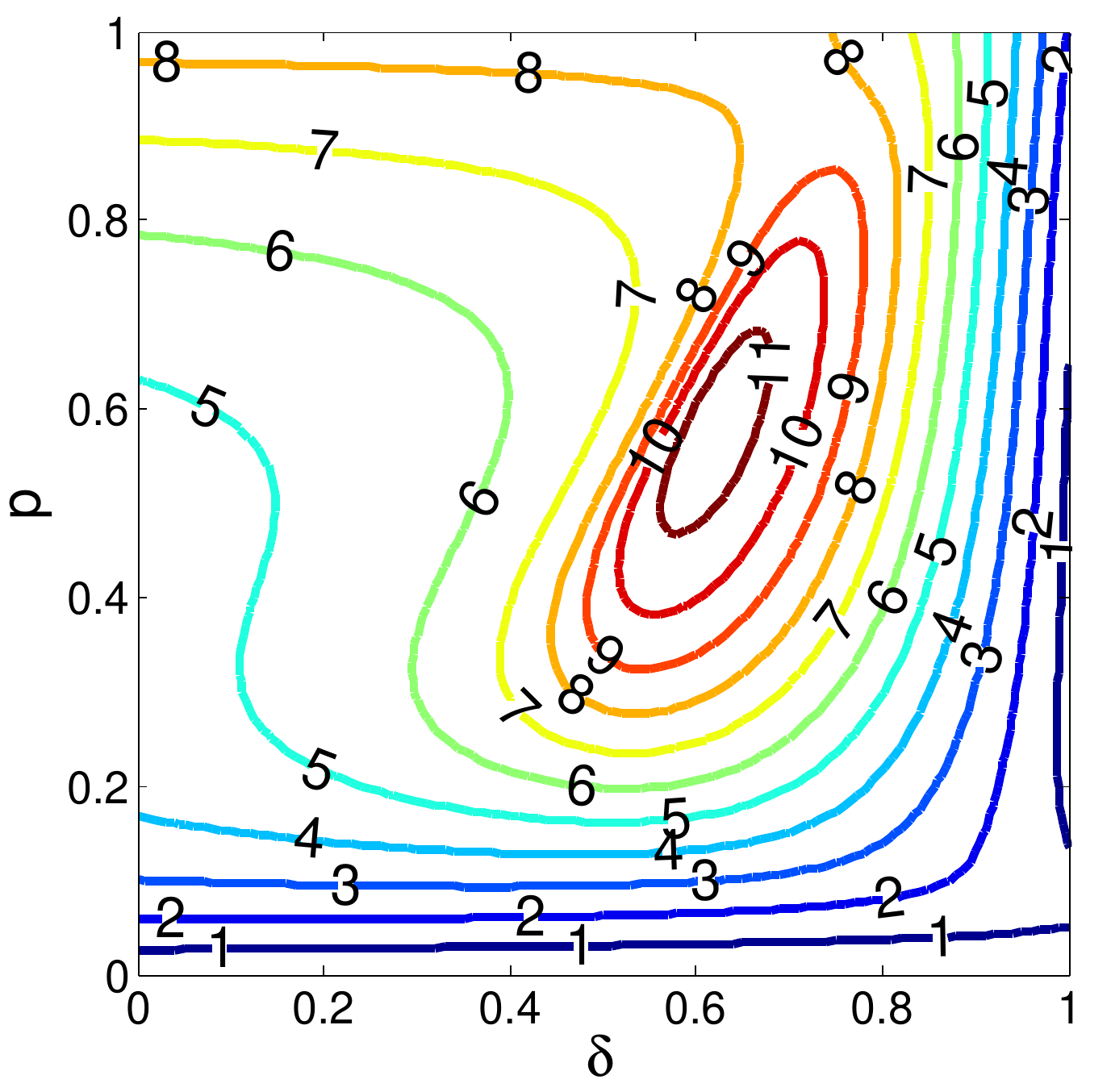}
  \caption{Operator's Revenue}\label{fig:Revenue}
\end{minipage}
\vspace{-5mm}
\end{figure*}

\section{Simulation Results}\label{sec:simu}


In this section, we numerically study how the network access valuation parameter $\rho_i$ and the mobility pattern $\boldsymbol{\eta}_i$ influence subscriber $i$'s membership selection decision, given other system parameters fixed.

In Section \ref{subsec:small}, we will first simulate a small network with $2$ APs and $1$ Alien, to gain insights of a single user's best choice.
Then, in Section \ref{subsec:large}, we simulate a large network with $100$ APs and $10$ Aliens to understand the system-level performance.

\subsection{A Small Network Example}\label{subsec:small}

We simulate a small network with $2$ subscribers (each owns an AP) and $1$ Alien.
We study how subscriber $1$'s network access valuation parameter $\rho_1$ and his probability of staying at home $\eta_{1,1}$ affect his membership selection.

We assume that the revenue sharing ratio $\delta=0.5$. The price at both APs is the same $p=1$.
The mobility patterns of subscriber $2$ and the Alien are same: $\boldsymbol{\eta}_2=\boldsymbol{\eta}_a=[1/3,1/3,1/3]$.
We assume that $\rho_1 \in [0,1]$. Subscriber $1$ stays at home with probability $\eta_{1,1}$, and travels to AP $2$ and outside the Wi-Fi coverage with a same probability $\eta_{1,2}=\eta_{1,0}=(1-\eta_{1,1})/2$.

Figure \ref{fig:Parameter} shows subscriber $1$'s membership selection decision in the equilibrium (Definition \ref{NEImixed} in Section \ref{subsec:mexed}), under different $\rho_i \in [0,1]$ and $\eta_{1,1} \in [0,1]$.
The color represents the value of $\alpha_1$, which is subscriber $1$'s probability of choosing to be a Bill.
The black region corresponds to $\alpha_1=1$, and the white region corresponds to $\alpha_1=0$.
The color in between corresponds to a mixed strategy of $\alpha_1\in (0,1)$, as shown in the colorbar on the right.

Figure \ref{fig:Parameter} shows that when $\eta_{1,1}$ is large enough (i.e., larger than $0.82$), i.e., subscriber $1$ stays at home most of the time, his will always choose to be a Bill with the probability $\alpha_1=1$, independent of subscriber $2$'s membership decision.
As $\eta_{1,1}$ becomes smaller and $\rho_1$ becomes larger, the performance and payment during roaming becomes increasingly important, so subscriber $1$ starts to choose a mixed strategy with a smaller number of $\alpha_1$.
When $\eta_{1,1}$ is small enough and $\rho_1$ is large enough, e.g., the right bottom corner of Figure \ref{fig:Parameter}, he will always choose to be a Linus with a probability $1-\alpha_1=1$.




\subsection{Simulation Results for Large Network}\label{subsec:large}

In this subsection, we simulate a larger network with $100$ APs and $10$ Aliens.

\subsubsection{Impact of Location Popularity}

We first study how the location popularity of an AP affects the subscriber's membership selection decision.
The location popularity is measured by the probabilities of users showing up at that location.
For simplicity, we assume that all users show up at the same location with the same probability, and the location popularity of APs $1$ to $100$ increases.

Figure \ref{fig:LP} shows each of the $100$ subscribers' membership selection decision.
The three curves represent three different values of the subscriber's network access valuation parameter $\rho_i$.
Under a given $\rho_i$, the subscriber's probability of choosing to be a Bill increases with his AP location popularity.
The reason is that a subscriber whose AP is located at a more popular location can earn more revenue from other Bills and Aliens.
For a particular subscriber (a fixed AP index), as his network access valuation $\rho_i$ increases, his probability of choosing to be a Bill decreases.
This is because he cares more about the network access benefit when roaming, and hence is more willing to be a Linus to enjoy free access and consume more data during roaming.

%

\subsubsection{Operator's Revenue}

Finally we discuss how the operator can utilize the analysis in this paper to optimize its revenue.
In particular, the operator can optimize $p$ and $\delta$, based on the users' equilibrium membership selection and network access decisions.

Figure \ref{fig:Revenue} presents contours of the operator's revenue with respect to $p$ and $\delta$.
In this case, the optimal price and revenue sharing ratio for the operator are $p^* = 0.58$ and $\delta^* = 0.63$, which lead to an average revenue of $11.40$ (per time slot) for the operator.
We can further see that the operator's maximum revenue is approximately $8.00$ if there is no incentive (i.e., $\delta = 0$), in which case the optimal price is around $p=1$.
On the other hand, the operator's maximum revenue is approximately $1.00$ under another extreme case with the maximum incentive (i.e., $\delta =1$), where the optimal price is around $p=0.3$.
Hence, with proper incentive, the operator can increase its maximum revenue up to $30\%$ and $90\%$, respectively, compered with those cases without incentive and with maximum incentive.

%

\section{Conclusion}\label{sec:conc}

In this paper, we set up a two-stage membership selection and network access game model for the crowdsourced wireless community network.
We analyze the game equilibria of both games systematically, and show that such an equilibrium analysis can help the operator make optimal pricing and incentive mechanism design.
We show that a user with a more popular home location, a smaller probability of travelling, or a smaller network access evaluation is more likely to choose to be a Bill.
There are several interesting directions for future researches.
First, it is interesting to theoretically study the operator's optimal pricing design.
Our result in this work serves as an important first step towards this problem.
Second, it is also interesting to study the problem under incomplete information, where some parameters (e.g., user network access valuation $\rho$, \rev{user mobility pattern $\boldsymbol{\eta}$}) are private information.

\end{document}